\documentclass{emulateapj}

\usepackage{graphicx}
\usepackage[hypertex]{hyperref}

\def \nh {${\rm N_{H}}$}
\def \eg {e.g.}
\def \ie {i.e.}
\def\spose#1{\hbox to 0pt{#1\hss}}
\def\ltsim{$\mathrel{\spose{\lower 3pt\hbox{$\sim$}}
        \raise 2.0pt\hbox{$<$}}$\thinspace}
\def\gtsim{$\mathrel{\spose{\lower 3pt\hbox{$\sim$}}
        \raise 2.0pt\hbox{$>$}}$\thinspace}
\newcommand{\thin }{\thinspace}

\newcommand{\msun }{${\rm M_{\odot}}$}
\newcommand{\lsun }{${\rm L_{\odot}}$}

\newcommand{\src }{NGC\thinspace 4649}
\newcommand{\zfe }{${\rm Z_{Fe}}$}

\newcommand{\chandra }{{\em Chandra}}
\newcommand{\genx}{{\em Generation-X}}

\newcommand{\xspec }{{\em Xspec}}

\newcommand{\ciao }{{\em CIAO}}
\newcommand{\caldb }{{\em Caldb}}
\newcommand{\heasoft }{{\em Heasoft}}
\newcommand{\ned}{{\em{NED}}}

\newcommand{\xmm }{{\em XMM}}

\newcommand{\mbh} {${\rm M_{BH}}$}

\newcommand{\lj }{${\rm L_J}$}

\newcommand{\mbul}{${\rm M_{bul}}$}

\newcommand{\mpl}{$M_\odot L_\odot^{-1}$}

\newcommand{\dtwentyfive}{${\rm D_{25}}$}
\newcommand{\twomass}{2MASS}
\newcommand{\sigmac}{$\sigma_*$}

\slugcomment{Accepted for publication in the Astrophysical Journal}

\begin{document}
\title{Weighing the quiescent central black hole in an elliptical galaxy with X-ray emitting gas}
\author{Philip J. Humphrey\altaffilmark{1}, David A. Buote\altaffilmark{1},
Fabrizio Brighenti\altaffilmark{2,3}, Karl Gebhardt\altaffilmark{4} and William G. Mathews\altaffilmark{3}}
\altaffiltext{1}{Department of Physics and Astronomy, University of California at Irvine, 4129 Frederick Reines Hall, Irvine, CA 92697-4575}
\altaffiltext{2}{Dipartimento di Astronomia, Universit\`{a} di Bologna, Via Ranzani 1, Bologna 40127, Italy}\altaffiltext{3}{University of California Observatories, Lick Observatory, University of California at Santa Cruz, Santa Cruz, CA 95064}
\altaffiltext{4}{Astronomy Department, University of Texas, Austin, TX 78712}
\begin{abstract}
We present a \chandra\ study of the hot ISM in the giant elliptical galaxy \src.
In common with other group-centred
ellipticals, its temperature profile rises with radius in the outer parts of the galaxy,
from $\sim$0.7~keV at 2~kpc to $\sim$0.9~keV by 20~kpc. However,
within the central $\sim$2~kpc the trend reverses
and the temperature peaks at $\sim$1.1~keV within the innermost 200~pc.
Under the assumption of hydrostatic equilibrium, we demonstrate that the central temperature spike arises 
due to the gravitational influence of a quiescent central super-massive black hole.
We constrain the black hole mass (\mbh) to 
$(3.35^{+0.67}_{-0.95})\times 10^9$\msun\ (90\% confidence), in good agreement with stellar kinematics 
measurements. This is the first direct measurement of \mbh\ based on studies of hydrostatic
X-ray emitting gas, which are sensitive to the most massive black holes,
and is a crucial validation of both mass-determination
techniques. This agreement clearly demonstrates the gas must be close to hydrostatic,
even in the very centre of the galaxy, which is consistent with the lack of morphological
disturbances in the X-ray image. 
\src\ is now one of only a handful of galaxies for which \mbh\ has
been measured by more than one method.
At larger radii, we were able to decompose the gravitating mass profile into stellar and dark
matter (DM) components. Unless one accounts for the DM, a standard Virial analysis of the stars
dramatically over-estimates the stellar mass of the galaxy.
We find the measured J-band stellar mass-to-light ratio,
$1.37\pm0.10 M_\odot L_\odot^{-1}$, is in good agreement with simple stellar population 
model calculations for this object.
\end{abstract}

\section{Introduction} \label{sect_introduction}
It is becoming increasingly clear that supermassive black holes (SMBHs) in the centres of 
galaxies are intimately involved in the evolution of their hosts. 
Models of galaxy formation suggest that energy injection from
active galactic nucleus (AGN) outbursts, in particular during hierarchical assembly, can have a significant
impact on the structure of the evolving galaxy \citep[\eg][]{silk98a,dimatteo05a}.
In galaxy groups and clusters, cavities in the hot intra-cluster medium (ICM)
 are a direct, observable manifestation of an episodic,
AGN-driven feedback process acting on the ICM \citep[\eg][]{birzan04a}. Such feedback 
has been implicated in global heating of the gas \citep[\eg][]{donahue06a}, 
redistributing metals through the ICM \citep[\eg][]{mathews04a} and 
invoked as a possible solution to the ``cooling paradox'' \citep[][and references therein]{mathews03a}.

Evidence for SMBHs has been found in increasingly large numbers of 
non-active galactic nuclei, leading to
the picture of ubiquitous SMBHs in 
stellar spheroids which has begun to
emerge \citep[][for a review]{ferrarese05a}.
The black hole mass (\mbh) has been found to correlate with properties of the host,
such as luminosity or mass of the bulge \citep{kormendy95a,magorrian98a,mclure02a,marconi03a},
the central light concentration \citep{graham01a},
or the central velocity dispersion (\sigmac) of the galaxy 
\citep{gebhardt00b,ferrarese00b,tremaine02a}. 
These observational relationships are starting to become a key constraint on galaxy-formation models 
\citep[\eg][]{robertson06a,granato04a}. There remain however, concerns over
the exact shape of the relations. In particular, different authors have found
different slopes for the  \mbh-\sigmac\ relation \citep{tremaine02a,ferrarese05a},
and there are suggestions it may deviate from a simple powerlaw relation
\citep{wyithe06a}. Possible disagreement in the masses of the largest SMBHs 
inferred from the 
\mbh-\sigmac\ and \mbh-bulge luminosity relations \citep{lauer07a} 
raise questions over which relation is more fundamental.
Clearly, further progress requires a careful assessment of possible systematic biases 
in the black hole mass measurements, especially in the high-mass regime.

A few independent techniques have evolved to measure the masses of SMBHs,
each relying on its own set of simplifying assumptions.
In galaxies with quiescent nuclei, the presence of the black hole is measured by its 
gravitational 
influence on the kinematics of either the stars or, if present, a central ionized gas
disk \citep[\eg][]{gebhardt03a,macchetto97a}. Since gas disks are not found in every
galaxy, stellar kinematical studies, based on measuring the line-of-sight velocity
dispersion profile are more generally applicable. Such studies are, however,
complicated by a strong degeneracy between orbital structure and the gravitating
mass profile, particularly in slowly-rotating galaxies \citep{binney82}. 
Although measuring high-order moments of the 
velocity dispersion profile allows the degeneracy to be partially broken \citep{vandermarel93a},
the solution is not, in general, unique \citep{valluri04a}. To obtain interesting
constraints on the SMBH mass, therefore, additional constraints such as a restricted
range of allowed orbits or smoothly-varying phase-space are required 
\citep[\eg][]{magorrian98a,gebhardt03a}. If these assumptions are unrepresentative,
however, \citeauthor{valluri04a} pointed out that the systematic errors in the 
\mbh\ determination
can exceed the statistical errors. 

Given the good agreement in the 
shape of the \mbh-\sigmac\ relation measured using masses obtained with different
methods \citep[\eg][]{gebhardt00c,ferrarese01a}, it seems unlikely that SMBH masses
determined from state-of-the-art ``3-integral'' methods \citep{vandermarel98a,gebhardt03a}
are {\em on average} in error. Nevertheless, a proper assessment of the systematics inherent
in the mass determination, by comparing masses {\em for the same black hole} measured
using different techniques is of considerable importance. 
To date there are only a handful of galaxies for which recent stellar kinematical 
\mbh\ measurements have been compared with at least one reliable, independent technique.
Stellar and gas dynamical estimates of \mbh\ have been found to agree reasonably in at
least two systems, NGC\thin 3379 and Cen\thin A \citep{shapiro06a,marconi06a}, although
there is evidence of non-circular motions in NGC\thin 3379, which can lead to a systematically
mis-estimated \mbh. Similar agreement has been found for M\thin 81 \citep{kormendy04a}, 
although the stellar kinematics measurement was based on ``2-integral'' models,
which are potentially unreliable as they place constraints on the orbital structure which appear
unrealistic, such as assuming the orbits are described entirely by only two integrals of motion
\citep[for more discussion see][]{ferrarese05a}.
Non-circular gas motions may explain strong discrepancies
between \mbh\ measured by these different techniques in the galaxies IC\thin 1459 and NGC\thin 4335
\citep{cappellari02a,verdoeskleijn02a}. Despite this concern, the \mbh-\sigmac\ relation is usually
constructed by combining samples of galaxies in which \mbh\ has been determined
from stellar kinematics and samples for which gas dynamics has been used \citep[\eg][]{tremaine02a},
and there is little evidence that 
the sub-samples obey appreciably different relations.

Other methods of determining \mbh\ are available in galaxies containing active galactic 
nuclei (AGNs), such as measuring the motions of masing gas \citep{miyoshi95a} or
``reverberation mapping'', using the variability of the  AGN \citep[\eg][]{peterson04a}. 
In two cases \mbh\ determined from reverberation mapping has been compared to stellar 
kinematics results \citep{davies06a,onken07a}, but since the uncertainties in the former 
technique lead to masses accurate only to within a factor of $\sim$3 \citep{onken04a}, these
measurements do
not provide good constraints on the validity of the stellar models.

A new method to identify SMBHs in giant elliptical galaxies was suggested by 
\citet{brighenti99c}, who pointed out that the gravitational influence of a 
central black hole should have an impact on the distribution of the characteristically
hot interstellar medium (ISM). The ISM should gradually flow inwards as it loses energy
from the emission of thermal X-rays but, since the local cooling time far exceeds the 
local free-fall timescale, the flow is highly subsonic and the gas remains close to hydrostatic equilibrium
\citep{mathews03a}. In the powerful gravitational field close to the black hole, the inflowing
gas should be sufficiently compressed to cause a strong temperature peak, which could be used
as a black hole diagnostic.
However, detections of central temperature peaks in galaxies are rare \citep{humphrey04b,humphrey05a,humphrey06a}
and this hypothesis has, therefore, yet to be verified observationally. In part this reflects the 
difficulty in obtaining precise temperature constraints from X-ray spectroscopy on scales 
of a few hundred parsecs. AGN-induced disturbances are also common in the
centres of early-type galaxies \citep{jones02a,birzan04a} and these may disturb the gas 
enough to destroy many such hot cores.

In this paper, we present a detailed \chandra\ study of \src,
a nearby, very relaxed and X-ray bright galaxy which is arguably
the most promising candidate in which to search for this effect.
Based on stellar kinematical studies, 
\citet{gebhardt03a} reported a black hole mass of $(2.0^{+0.4}_{-0.6})\times 10^9$\msun\ 
(1-$\sigma$ errors), placing it at the upper end of the 
\mbh-\sigmac\ relation.
A modest archival \chandra\ dataset has already revealed a remarkably relaxed
X-ray morphology \citep[implying hydrostatic equilibrium is a good approximation:][]{buote95a} 
and a temperature profile which peaks in the centre \citep{humphrey06a,randall03}. The object
is faint in the radio \citep{condon02a}, indicating that it does not host a powerful AGN, and
there is little evidence of morphological disturbances indicating
AGN-ISM interaction (\S~\ref{sect_image}). On a large scale, excellent agreement between
the mass determined from globular cluster kinematics and X-ray methods \citep{bridges06a}
also provides good support for hydrostatic equilibrium.
We present deep new data which enable us to obtain interesting temperature constraints on
scales as small as $\sim$150~pc, which is crucial given the $\sim$60~pc sphere of 
influence of the SMBH on the gas\footnote{Defined as $GM_{BH}/c_s^2$, where $c_s$ is the central gas
sound-speed (in the absence of a black hole), $\sim 490 km\ s^{-1}$ from our data. This is slightly smaller than the sphere of 
influence on the stars, $GM_{BH}/\sigma_*^2$,
which is $\sim$100~pc for the measured $\sigma_*=385 km\ s^{-1}$ \citep{gebhardt03a}.}.
For the first time in {\em any} system we are able to place constraints on 
\mbh\ from hydrostatic X-ray gas.

We adopted a nominal distance of 15.6~Mpc for \src, based on the I-band SBF distance
modulus estimate of \citet{tonry01}, which was corrected by -0.16 to account for recent revisions to
the Cepheid zero-point. At this distance, 1\arcsec\ corresponds to 76~pc. 
Unless otherwise stated, all error-bars represent 90\%\ confidence limits.

\section{X-ray data analysis}
\begin{figure*}
\centering
\includegraphics[width=7in]{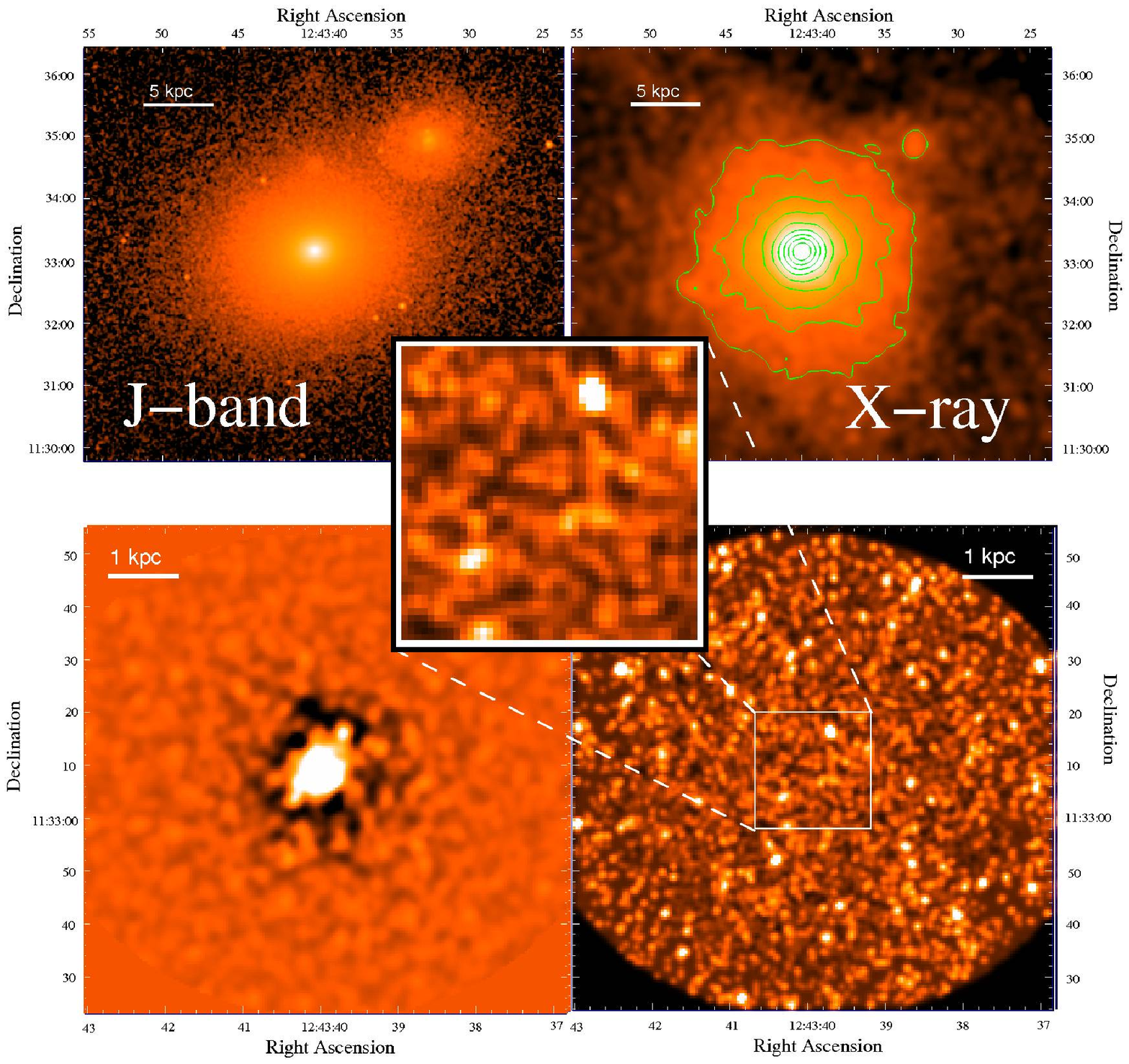}
\caption{{\em Top left:} J-band \twomass\ image of \src. Note the interloper spiral galaxy NGC\thin 4647
to the northwest. {\em Top Right:}  point-source cleaned \chandra\ 
image of \src, smoothed by a 5\arcsec\ Gaussian kernel, shown with arbitary, logarithmically-spaced
contours. Note there is only a small amount of emission apparently associated with NGC\thin 4647. 
{\em Bottom left:}  ``Unsharp-masked'' \chandra\ image of the central part of \src\
({\bf note the different scale}), created by subtracting an image 
smoothed with a 3\arcsec\ Gaussian kernel from one smoothed by a 1\arcsec\ kernel. The dark ring
around the central, bright emission is an artefact of the smoothing process. {\em Bottom right:}
``residual significance'' image (see text) of the central part of the galaxy, indicating deviations from a model
derived from elliptical isophote fitting to the data. The galaxy is very relaxed at all scales, and we
do not confirm recent suggestions of morphological disturbances. {\em Center:} 
central 22\arcsec$\times$22\arcsec\ region of the residual significance image.} \label{fig_image}
\end{figure*}

\begin{deluxetable}{llr}
\tablecaption{Observation summary\label{table_observations}}
\tablehead{
\colhead{ObsID} & \colhead{Start Date} & \colhead{Exposure (ks)} 
}
\tablewidth{3in}
\startdata
785 & 2000 Apr 20 & 18 \\
8182 & 2007 Jan 31 & 48 \\
8507 & 2007 Feb 1 & 15\\
{\bf Total} & \ldots & {\bf 81}\\
\enddata
\tablecomments{Details of the \chandra\ observations used in the present analysis. For each 
dataset we quote the observation ID number (ObsID), the start date and the exposure time, having
removed periods of background ``flaring'' (Exposure)}
\end{deluxetable}

The region of sky containing \src\ has been imaged by the \chandra\ ACIS instrument
in ACIS-S configuration on three separate occasions, 
as listed in Table~\ref{table_observations}. We processed each dataset independently,
using  the
\ciao\ 3.4 and \heasoft\ 5.3.1 software
suites, in conjunction with \chandra\ calibration database (\caldb)
version 3.3.0.1. To ensure up-to-date calibration, all data were
reprocessed from the ``level 1'' events files, following the standard
\chandra\ data-reduction threads\footnote{{http://cxc.harvard.edu/ciao/threads/index.html}}.
We applied the standard correction to take account of the time-dependent gain-drift
and charge transfer inefficiency,
as implemented in the \ciao\ tools. To identify periods of enhanced
background (``flaring''), which seriously degrades the signal-to-noise (S/N)
we accumulated background lightcurves for each dataset from
low surface-brightness regions of the active chips, excluding obvious
point-sources. Periods of flaring were identified by eye and
excised. The final exposure times are listed in Table~\ref{table_observations}. 

\subsection{X-ray image} \label{sect_image}
For each
dataset we generated a full resolution image in the 0.3--7.0~keV energy-band 
and a corresponding
exposure map computed at an energy of 1.7~keV. 
Point sources were detected in each image with the \ciao\ {\tt wavdetect} task,
which was set to search for structure at scales of 1, 2, 4, 8 and 16
pixels, and supplied with the exposure-maps to minimize spurious detections at the image boundaries.
The detection threshold was set to $10^{-6}$, corresponding to \ltsim 1 spurious source
detections per chip. To improve the signal-to-noise (S/N) of the data, we combined the 
resulting  images and exposure maps. 
Given errors in the absolute astrometry of \chandra, we corrected the aspect onto a 
common frame of reference before addition. To achieve this registration, we 
matched common point-sources in each image and shifted the data to ensure the maximum
number of such sources overlapped. Since there can be variability in a significant fraction of the 
LMXBs in any galaxy (or, indeed, in background AGN) we did not require all point sources
to match. Having generated a merged image and exposure-map, to improve signal-to-noise (S/N),
we re-detected the point sources. All detected sources were confirmed by visual inspection, and, for each,
appropriate elliptical regions containing approximately 99\%\ of its photons were generated.

We examined the image for evidence of 
morphological disturbances which may suggest deviations from 
hydrostatic equilibrium. First we  processed each image to remove the point-sources.
To do this, we replaced all photons within the source detection ellipse of each source with artificial 
data, using the algorithm described in detail in \citet{fang08a}. 
In short, since the {\tt wavdetect} task can be used to generate a smoothed, ``normalized'' 
source-free image, we replaced the data within each source detection ellipse from our real image
with data from the corresponding pixels on the source-free image, after adding Poisson noise
and taking exposure-map variations into account. We found this procedure generally worked extremely
well at removing sources. The data were then flat-fielded with the exposure map and examined by eye. 
We show in Fig~\ref{fig_image}  the image of \src, having smoothed it slightly
with a 5\arcsec\ (0.4~kpc) Gaussian kernel. The image is remarkably round and shows no
evidence of disturbances or structure. It is interesting to note that the X-ray image is
rounder than the optical light (which is, nonetheless, quite round, with a major/ minor
axis ratio of \ltsim 1.2). This is unsurprising since gas in
hydrostatic equilibrium traces the {\em potential} 
\citep[the ``X-ray shape theorem'':][]{buote94}, which is rounder than the mass distribution giving
rise to it. 

To seek evidence of more subtle structure, we experimented with 
``unsharp masking'' the image, \ie\ differencing the image smoothed at two different scales.
We did not find convincing evidence of any morphological disturbances for any of the scales
(using smoothing kernels from  3--20\arcsec) we tried. We show in Fig~\ref{fig_image} a
typical ``unsharp-masked'' image, illustrating the lack of substructure (note that the dark ring around the
central bright region is an artefact of the smoothing process).
\citet{shurkin07a} recently 
argued for the presence of AGN-induced cavities in the centre of \src, but, using deeper data
than these authors, we found little evidence for such features. Performing the ``unsharp masking''
they describe, we found little evidence for the supposed cavities, save features which are practically
indistinguishable from 
erratic  structure arising as an artefact of the image processing. Nonetheless, to assess whether
any such depressions do exist, we first fitted elliptical isophotes to the central part of the X-ray image
with the {\em PC-IRAF} task {\tt ellipse}. We then extracted the total counts both in the image and
the model within two circular apertures of 4\arcsec\ diameter placed approximately at the positions reported 
by \citet{shurkin07a} for the depressions. One can then compute  $\chi^2=$(data-model)$^2$/model which can
be used to assess the significance of the features. A standard $\chi^2$ would assume that we 
had performed a random trial, however, which is clearly incorrect here. Instead, we performed
Monte-Carlo simulations to assess the true distribution of $\chi^2$. We performed 1000 such simulations, 
in each of which an artificial image was generated by
adding Poisson noise to the model image. The model was then subtracted and the resulting image smoothed 
(with a 2\arcsec\ Gaussian) to bring out low counts structure at the scale of interest 
which lay close to the apparent north and
south radio lobes (by which we mean within 25\arcsec$\times$25\arcsec\ boxes centred on each lobe). 
We repositioned the circular apertures at the corresponding minima and computed 
$\chi^2$ from the (unsmoothed) artificial data. We found that 5.3\%\ of the simulations gave $\chi^2$ larger than
we obtained with 
the real data, implying that the putative cavities are significant only at \ltsim 2-$\sigma$. We therefore
consider the evidence for AGN-induced cavities in the hot gas to be weak. 

To investigate the presence of other possible structures, we 
show in Fig~\ref{fig_image} an indication of deviations
from our model fit by plotting (data-model)$^2$/model, which corresponds to the 
$\chi^2$ residual in each pixel. To bring out structure, we smoothed this image with a Gaussian kernel
of width 0.5\arcsec\ (1 pixel). As is clear from this ``residuals significance'' image, there is 
no evidence of any coherent residual features (with the possible exception of one or two 
point-sources that may have been imperfectly removed). 
The X-ray emission is therefore very smooth and symmetric.

 \citet{randall05a,randall03} adaptively smoothed \chandra\ and \xmm\ images of \src\ (using 
shallower \chandra\ data than the present work) and
identified radial features they dubbed as ``fingers of emission''. However, although adaptive 
smoothing can give a useful overall impression of the data, it tends to
introduce artificial structure into images. 
Our data are deeper and we do not find convincing evidence of such features.
Furthermore, visual comparison of the authors' \chandra\ and \xmm\ images suggests
some differences; in particular the radially-averaged, azimuthal surface-brightness plots 
do not obviously resemble each other, with the \xmm\ data appearing (by eye) to show a larger
amplitude variation with position angle than the \chandra\ data. Since \xmm\ is less able to resolve out LMXBs,
we speculate that unresolved point-sources may be largely responsible for the significant
structure reported for the \xmm\ data.
Even taking the supposed \xmm\ structures at face value, they would imply
azimuthal variations in gas density of only$\sim$5\%. 
Small variations of this magnitude do not constitute significant
deviations from hydrostatic equilibrium and are smaller, in general, than 
our measurement errors. Based on the shallower observation, \citet{randall03} also report 
a $\sim 2.5$-$\sigma$ detection of 
a central $\sim$5\arcsec\ structure, but inspection of the ``residual significance'' image shown in 
Fig~\ref{fig_image} indicates there are no significant residual structures in the central region.


\subsection{Spectral analysis} \label{sect_spectra}
We extracted spectra in a series of concentric, contiguous annuli, placed at the X-ray centroid. 
We determined
the centroid iteratively by placing a 0.5\arcmin\ radius aperture at the nominal
galaxy position (obtained from \ned) and computing the X-ray centroid
within it. The aperture was moved to the newly-computed centroid, and the
procedure repeated until the computed position converged.
The widths of the annuli were chosen so as to contain approximately the same number of 
background-subtracted photons, and ensure there were sufficient photons to perform useful
spectral-fitting (in this case we had \gtsim 2000 background-subtracted photons per spectrum). 
We placed a lower limit of 2.5\arcsec\ on the annulus width, to ensure
that the instrumental spatial resolution  does not lead to strong mixing between the spectra in 
adjacent annuli. The data in the vicinity of any
detected point source were excluded, as were the data from the vicinity
of chip gaps, where the instrumental response may be uncertain. 
To prevent possible low-level contamination from any hot gas in the central part of the 
interloper galaxy NGC\thin 4647, we additionally excluded data in its vicinity from the 
9.3--13.1~kpc annulus, which intersects its centre. We extracted
products from all active chips, excluding the S4 (which suffers from
considerable ``streaking'' noise). 
Appropriate count-weighted spectral response
matrices were generated for each annulus
using the standard \ciao\ tasks {\tt mkwarf} and {\tt mkacisrmf}. For each spectrum,
we estimated the background using the method outlined in \citet{humphrey06a}.
We extracted identical products individually for each dataset,
taking into account the astrometric offsets between them (determined from
the image registration discussed above). The source and background spectra were
added, and the response matrices averaged using the standard \heasoft\ tasks
{\tt addrmf} and {\tt addarf}, with weights based on the number of photons in
each spectrum.  We found there was very little emission outside $\sim$5\arcmin\
(23~kpc), indicating truncation of the X-ray halo outside this radius as
\src\ falls into the Virgo cluster. 

Spectral-fitting was carried out in the energy-band 0.5--7.0~keV.
 The spectra were rebinned to ensure a S/N ratio of at least
3 and a minimum of 20 photons per bin (to validate $\chi^2$ fitting).
We fitted data from all annuli simultaneously using
\xspec. To model the hot gas we adopted a {\bf vapec} component, plus a 
thermal bremsstrahlung component to account for  undetected point-sources 
\citep[this model gives a good fit to the composite spectrum of the detected sources in nearby
galaxies:][]{irwin03a}. In common with our previous analysis \citep{humphrey05a,humphrey06a}
we adopted a slightly modified form of the \xspec\ {\tt vapec} code to enable us to 
tie the ratio of each elemental abundance with respect to Fe between each annulus, 
while the Fe abundance (\zfe) was allowed to fit freely. 
To improve S/N \zfe\ was tied
between adjacent annuli where constraints in individual annuli were poor.  
We allowed the global ratios of O, Ne, Mg, Si, S, and Ni with respect to Fe to fit freely,
and fixed the remaining ratios at the Solar value
\citep{asplund04a}.  The absorbing column density (\nh) 
was fixed at the Galactic value \citep{dickey90}. To account for projection effects, we used the 
{\bf projct} model implemented in \xspec. Since the X-ray emission appears truncated 
at large radius, we did not include a component to account for projection from 
outside our outermost annulus. However, as \src\ is within the Virgo cluster, we took account of
possible interloper cluster emission by 
including an additional hot gas component, with kT fixed at 2.5~keV \citep[\eg][]{gastaldello02a},
which was
assumed to have a constant surface brightness over the field of view. 
We obtained a reasonable fit ($\chi^2$/dof=1178/1088) to the spectra with this model, 
which is of comparable quality to our previous analysis of 7 systems \citep{humphrey06a}. 
The 
best-fitting abundances were in agreement with those reported for \src\ in that paper.
Error-bars were computed {\em via} 
the Monte-Carlo technique outlined in \citet{humphrey06a}, and we carried out 50 error simulations. 
The resulting temperature and density profiles are discussed below.

\section{Mass modelling} \label{sect_models}
\subsection{Entropy-Temperature method} \label{sect_entropy_model}
In our previous work \citep{humphrey06a,gastaldello07a,zappacosta06a} we considered a variety of different
approaches to solve the equation of hydrostatic equilibrium and obtain the mass 
profile. In our past analysis of \src, we adopted a parameterized model for the total
gravitating mass profile and an ad hoc model for the temperature profile, solving for the 
total gas density. Although this method allowed us to obtain interesting constraints, it 
relies to some extent on the adoption of a temperature profile with reasonable asymptotic
behaviour. We found a wide range of different temperature profiles for the galaxy-scale
objects in our sample, unlike the approximately ``universal'' temperature profiles of galaxy 
clusters and, to some extent, galaxy groups \citep{vikhlinin05a,gastaldello07a,rasmussen07a}.

As an alternative to these methods, in the present work we chose to exploit the fact
that the {\em entropy} profile is, in general, far smoother than the temperature or
density profiles and can typically be fitted by simple parameterized models
which are well-behaved asymptotically
\citep[\eg][]{donahue06a}. In what follows  we define the quantity
$S=\rho^{-2/3} kT/(\mu m_H)$, where $\rho$ is the gas density, T is the temperature, k is 
Boltzmann's constant, $m_H$ is the mass of a hydrogen atom and $\mu$ is the mean
molecular weight factor, taken to be 0.62. This is related to the actual entropy by a logarithm
and multiplicative constant and is proportional to the usual ``entropy'' proxy
used in the literature, $s=n_e^{-2/3} kT$ \citep[\eg][]{sanderson03a}. An advantage of considering the 
entropy profile is that stability against convection requires $dS/dr$ to be positive at 
all radii, if the hydrostatic equilibrium equation is valid, placing a further constraint
on meaningful models and improving our mass constraints. 

One can write the equation of hydrostatic equilibrium in terms of pressure, P, and
S as 
\begin{equation}
\frac{dx}{d r} = - \frac{2}{5}\frac{G M_{TOT}(r)}{r^2} S^{-3/5}
\end{equation}
where x=$P^{2/5}$, G is the universal gravitating constant, and $M_{TOT}$(r) is the total mass enclosed within
a given radius r. For a given distribution of S and $M_{TOT}$, this equation can be 
directly solved for x, provided the mass contribution of the gas can be neglected at 
all radii. Although this is true in the present work, since the gas contributes $<1$\%\ of the 
mass at all radii, in practice we allowed for a nonzero gas mass by differentiating 
and rearranging the equation
to give:
\begin{equation}
\frac{d}{dr}\left( r^2 S^{3/5}\frac{dx}{d r}  \right) +\frac{8G\pi r^2}{5 S^{3/5}}x^{3/2} = 
-\frac{2}{5}G\frac{dM}{dr} \label{eqn_hydro}
\end{equation}
where the mass enclosed, M, comprises only the dark matter, stellar mass and the black hole
mass. The gas mass is represented by the second term on the left hand side. 
For any parameterized distributions of S and M, this equation
can be solved uniquely for x using a Runge-Kutta method. This means that for any such parameterized
models of mass and entropy, the temperature and density profiles can be uniquely determined. 
We began all integrations at 
10~pc, assuming the gas contributes negligibly to the total mass within this radius. The starting value
for x (\ie\ central pressure) at 10~pc  was determined as a free parameter of our fit.


For each annulus in which we extracted data, we explicitly integrated appropriately weighted functions of 
pressure and temperature to compute the predicted contribution to the emission measure
(and hence mean density)
from gas in the corresponding shell (recalling that the data have been deprojected).  
We also computed an emission-weighted
temperature (ignoring the temperature and abundance dependence of the emissivity, since
neither term varies much over the field of view). We compared these averaged quantities to
our measured temperature and density data-points. See \citet[][Appendix B]{gastaldello07a}
for a detailed discussion of incorporating the plasma emissivity into the gas modelling.

\subsection{The stellar mass component} \label{sect_stars}
\begin{figure*}
\centering
\includegraphics[width=7in]{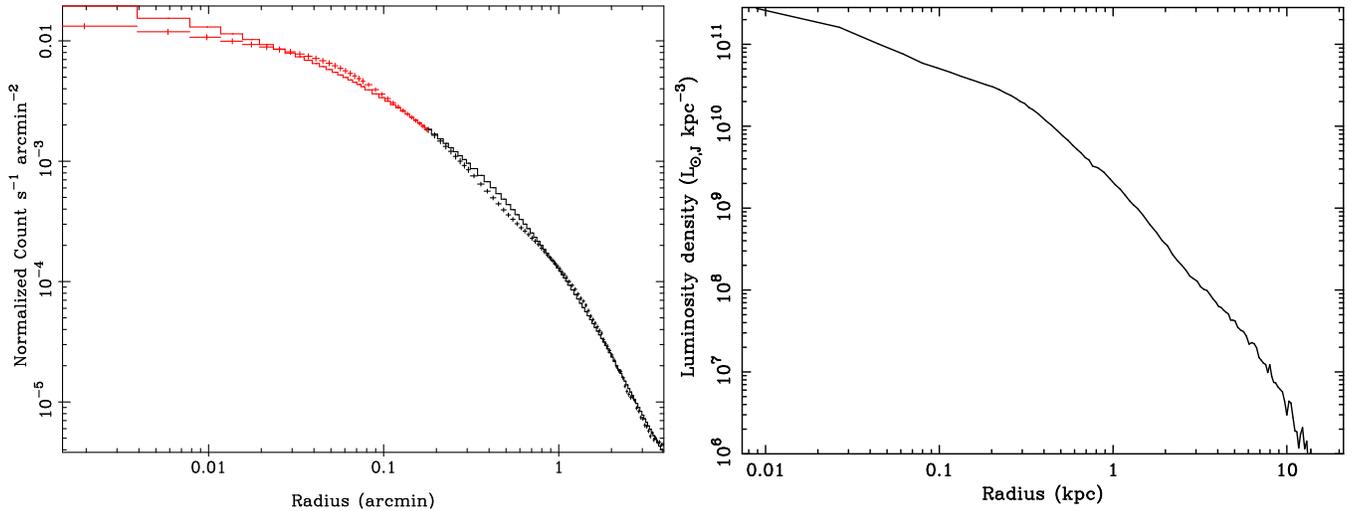}
\caption{{\em Left:} Surface brightness profile of the galaxy, in optical light,
shown with the best-fitting
Sersic model. The Sersic model describes the overall shape, but not the detail 
of the luminosity profile. To aid the fitting, a systematic error of 5\%\ has been 
added in quadrature with the statistical errors on each data-point. The
J-band \twomass\ profile is shown in black, and the renormalized I-band
HST data in red. {\em Right:} The non-parametrically
deprojected J-band luminosity density profile.} \label{fig_optical_light}
\end{figure*}
In order to place constraints on the black hole mass in the centre of 
a galaxy, it is imperative that we have an accurate picture of the stellar
mass in that region. To do this, we assumed that the 
stellar mass  follows the three dimensional distribution of the
optical light, provided we adopt data observed in as red a filter as 
possible. From our Lick index analysis \citep{humphrey06a}, we were able 
to estimate a mean age for the stellar population, as well as both a central
stellar abundance and an emission-weighted abundance for the galaxy as a whole.
Adopting the stellar population models of \citet{maraston05a}, the J-band 
M/\lj\ ratio should only change by $\sim$3\% depending on which of the two
abundances is assumed. This supports the assumption that mass follows light.
To ensure maximum accuracy at the scale of our innermost \chandra\
data-bins, we first considered the HST WFPC2 data with the F814W (approximately
I-band) filter \citep[for more details on the HST analysis see][]{humphrey07a}. 
We considered only data on the PC chip (to prevent the 
particular geometry of the WFPC2 plane complicating our analysis), limiting
our profile to a radius of $\sim$11\arcsec. To extend the coverage to larger
radii ($\sim$4\arcmin), we adopted the publicly-available \twomass\ 
J-band images, masking out all data within the \dtwentyfive\ ellipse 
\citep[as listed in][]{devaucouleurs91} of the interloper spiral galaxy 
NGC\thin 4647 (Fig~\ref{fig_image}) as 
well as from the vicinity of any bright point-sources.
Although we were not able exactly to match the filters used in both cases,
when the I-band HST profile is suitably rescaled, the profiles agree well 
outside $\sim$2\arcsec, implying that there is not a strong I-J colour
gradient over the HST field of view, and hence  we can use the HST data
as a proxy for a high-resolution J-band image.

By visual inspection, the axis ratio (major/ minor) of the optical isophotes 
is \ltsim 1.2 at all radii of interest (Fig~\ref{fig_image}), 
indicating that the galaxy is fairly round;
in what follows we adopt a spherical approximation. First we investigated
whether simple analytical models can adequately match the observed surface
brightness profile. We found that the de Vaucouleurs model significantly
over-estimated the optical light in the centre of the galaxy. A Sersic
profile with n$\simeq$2.6 was able to characterize the overall shape 
of the profile but, especially in the inner regions, could not capture
its fine detail (Fig~\ref{fig_optical_light}). 

Rather than using these simple presciptions, we instead adopted a 
numerical deprojection technique, which is justified given the 
high quality of the data. We divided the emission into a series of concentric shells with outer
and inner radii corresponding to the annuli in which our surface brightness
profile (Fig~\ref{fig_optical_light}) was computed. In each shell,
we assumed the  luminosity density  of the stars was constant, enabling
us to relate the observed surface brightness in each annulus to an 
appropriately weighted sum of emission from the shell it intersects and
those outside it \citep[\eg][]{kriss83a}. 
We assumed there was no emission outside the outermost shell, allowing us to 
compute the luminosity density (in \lsun\ kpc$^{-3}$) in the last shell directly from the 
measured surface brightness in our outer annulus. We then stepped inwards,
iteratively solving for the luminosity in each shell. This technique is effectively
the same as the ``onion peeling'' algorithm widely used to deproject spectra
\citep[\eg][]{buote00c}. We assumed a J-band solar magnitude of 3.71, for 
consistency with the stellar population models of \citet{maraston05a}. 
We have confirmed that the integrated light profile is in good agreement with the
V-band deprojected light profile used by \citet{gebhardt03a} (when scaled by a 
factor ~3.34 which, using \citeauthor{maraston05a}'s
models, accounts for the different photometric bands used given the 
metallicity and age of the stars in the centre of NGC\thin 4649). 
The profiles differed by less than 20\%\ over all scales at which they overlapped,
and typically agreed within a few percent (especially in the crucial inner kpc).
Such agreement provides further support for our assumption that the stellar M/\lj\ 
ratio is approximately constant with radius.

\section{Results} \label{sect_results}
\begin{deluxetable}{ll}
\tablecaption{The best-fit results\label{table_results}}
\tablehead{
\colhead{Parameter} & \colhead{Value} 
}
\startdata
$\chi^2$/dof & 18.3/20 \\
\mbh & $(3.35^{+0.67}_{-0.95})\times 10^9$\msun\  \\
stellar M/\lj & $1.37\pm 0.10 M_\odot L_\odot^{-1}$\\
$s_0$ & $2.00^{+0.26}_{-0.37}$ keV cm$^2$ \\
$s_1$ & $8.6^{+2.3}_{-1.4}$ keV cm$^2$ \\
$\beta_1$ & $1.5\pm0.18$ \\
$s_{brk}$ & $2.44^{+0.59}_{-0.40}$ kpc\\
$\beta_2$ & 1.07$\pm 0.05$
\enddata
\tablecomments{Selected best-fitting parameters (see text).}
\end{deluxetable}

\begin{figure}
\centering
\includegraphics[width=3.4in]{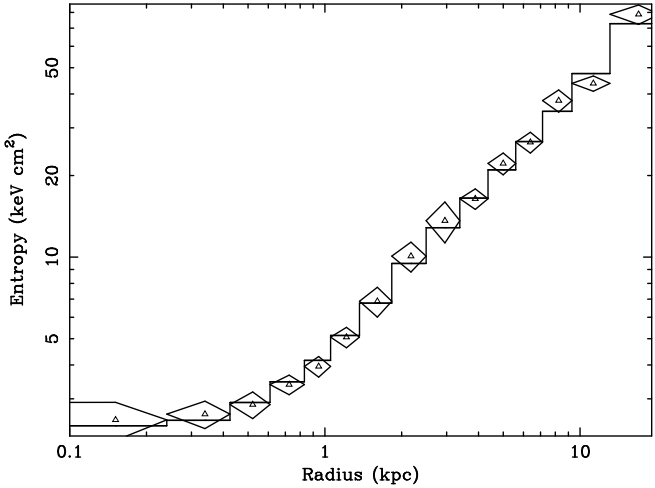}
\caption{The entropy profile of the data, shown along with the best-fitting model. 
Although the profile flattens at small radii
the entropy gradient is always positive, as expected for gas in hydrostatic equilibrium.
} \label{fig_entropy_profile}
\end{figure}

From our spectral-fitting results, we have determined mean temperature and 
density data-points within each annulus. We converted these data to a
mean (emission-weighted) entropy profile, as shown in Fig~\ref{fig_entropy_profile}.
We chose to fit this profile with a simple empirical model, for which we found 
a broken-powerlaw plus a constant model to be sufficient, \ie:
\begin{equation}
s  =  s_0 + s_1 f(r)
\end{equation}
where:
\begin{eqnarray}
f(r)  = &  \left( \frac{r}{s_{brk}} \right)^{\beta_1}  & (r < s_{brk}) \nonumber \\
 &  \left( \frac{r}{s_{brk}} \right)^{\beta_2}  & (s_{brk}\le r  ) 
\end{eqnarray}

As explained in \S~\ref{sect_models}, for a pair of parameterized models for entropy and 
the mass profile, one can uniquely determine the temperature (and density) profile. 
It is therefore possible to determine the best-fitting parameters of the mass model
by fitting the entropy and temperature data simultaneously, while varying the parameters
which describe the entropy and mass profiles. Since the fractional 
error-bars on the density are typically far larger than those on the temperature
(due to partial degeneracies between the continuum and the unresolved point-source component,
and between the gas density and abundance, for the range of temperatures considered here), we found this was
more constraining than fitting, for example, entropy and pressure at the same time.

To model the mass profile, in addition to the mass of the gas
we included a stellar mass component, as described in 
\S~\ref{sect_stars}, an NFW  dark matter model, following \citet{navarro97}, and a 
central supermassive black hole. We allowed the stellar mass-to-light (M/\lj) ratio, 
central black hole mass (\mbh), and total mass and concentration of the NFW dark halo
to fit freely.  We find that the global parameters of the dark matter (NFW) halo 
are highly degenerate with each other
without the application
of additional constraints \citep[\eg][]{humphrey06a}, but their exact values have little
bearing on the shape of the dark matter profile in the region of interest, so we will 
discuss them in detail in a future paper.

\begin{figure}
\centering
\includegraphics[width=3.4in]{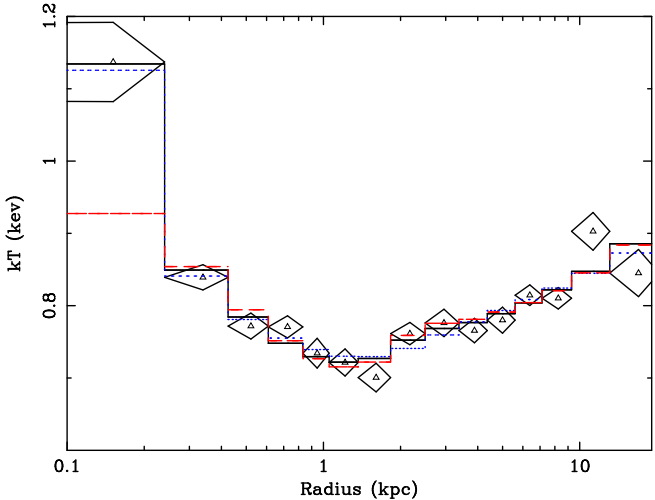}
\caption{Measured temperature profile of the hot gas in \src\ (error diamonds). 
Also shown are the best-fitting 
modelled profile from the entropy-temperature analysis (solid line; black) and the best-fitting
temperature profile if a central black hole is omitted (dashed line; red). In addition,
the dotted line (blue) indicates the best-fitting arbitrary parameterized model
(See \S~\ref{sect_nonparametric}) } \label{fig_temp_profile}
\end{figure}

\begin{figure}
\centering
\includegraphics[width=3.4in]{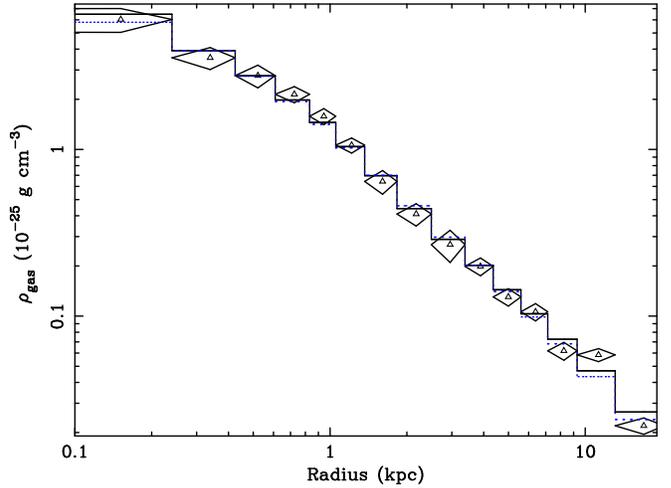}
\caption{Measured gas density profile of the hot gas in \src\ (error diamonds). 
Also shown are the best-fitting 
modelled profile from the entropy-temperature analysis (solid line; black) and the best-fitting
arbitrary parameterized model (dotted line; blue. See \S~\ref{sect_nonparametric}).} \label{fig_rho_profile}
\end{figure}

Because the measured entropy and temperature in each annulus were, by definition, 
averaged quantities, in contrast to the continuous, parameterized functions being fitted,
care was taken over comparing the model with the data. The model density profile (squared) was 
explicitly integrated over each bin to compute the emission measure, while the temperature
model was similarly integrated with $\rho_{gas}^2$ (emission) weighting.
 These quantities
correspond to the observables determined directly from the spectral fitting, and so 
could be used to compute the emission-weighted entropy (and temperature) for direct 
comparison with the data.
The best-fitting entropy model is shown in  Fig~\ref{fig_entropy_profile}, and exhibits
asymptotic (large radius) behaviour consistent with the expected $s\propto r^{1.1}$
from structure formation models incorporating purely gravitational processes \citep[\eg][]{tozzi01a}.
We also show the temperature and gas density profiles of the object, 
along with the best-fitting models in  Figs~\ref{fig_temp_profile} and Fig~\ref{fig_rho_profile}. 

\begin{figure}
\centering
\includegraphics[width=3.4in]{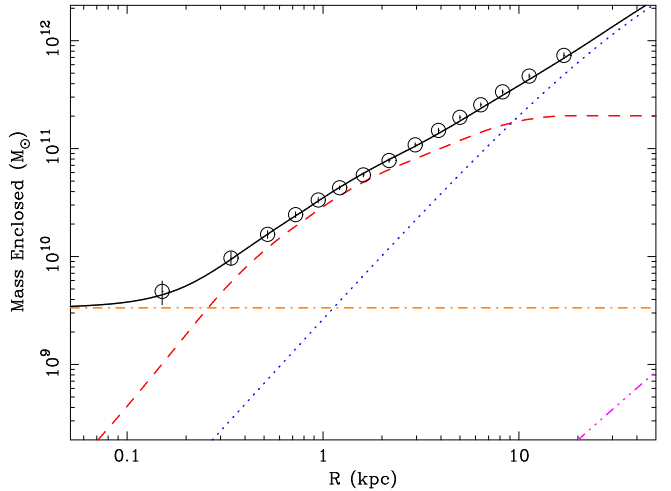}
\caption{Mass profile of \src\ determined from our fitting. The individual data-points
were derived from the traditional non-parametric method discussed in \citet{gastaldello07a}.
We also show our best-fitting mass-profile determined from our entropy-temperature modelling 
method (see text). We show separately the total mass enclose (solid line; black),
the stellar component (dashed line; red), the dark matter (dotted line; blue), the gas
(dash-dot-dot-dot line; magenta) and the black hole mass (dash-dot line; orange).
We stress that the models {\em do not} represent a fit of the data-points, which were 
derived independently.} \label{fig_mass_profile}
\end{figure}

The radial dependence of the best-fitting mass model is shown in Fig~\ref{fig_mass_profile},
along with the contribution of each of the separate mass components.
We find that the stellar mass component dominates the mass profile within $\sim$6~kpc, with a 
significant contribution of dark matter required to explain our observations at large radii. This 
is largely consistent with our previous analysis, despite the much simpler stellar mass profile 
we adopted in that work. Within our innermost annulus, however, we found that the rapidly-falling
stellar mass was unable to explain the strong peak in the temperature profile.
We show in Fig~\ref{fig_temp_profile} the best-fit temperature 
model with the black hole mass (\mbh) fixed at zero,
which clearly does not reproduce the sharp central spike and is only a marginally acceptable
fit (global $\chi^2$/dof=33.7/21). Allowing the black hole mass to fit 
freely enabled us to obtain an excellent fit to the data ($\chi^2$/dof=18.3/20),
an improvement significant at 99.94\%\ significance, based on an f-test. 
The best-fitting values of key fit parameters are shown in Table~\ref{table_results}.


\subsection{Traditional mass analysis} \label{sect_nonparametric}
In our analysis, we adopted a new approach to X-ray mass analysis, which we expect to give a 
more precise determination of \mbh\ than would be expected from other, more traditional 
approaches. However, it is also of interest to determine whether interesting results can be obtained
using the older methods, which have the advantage of less computational expense.
In order to investigate this, we additionally
analysed the data using a more orthodox mass-analysis approach (here dubbed the ``traditional method'').
In this case, we adopted ad hoc parameterized models for both the
temperature and density profiles. These models
were chosen to capture as much detail of the profiles as possible, but may not be a perfect
fit to the data. One can simply compute the mass enclosed within any
radius by  differentiating these models and inserting the derivatives into
an appropriately re-arranged form of the hydrostatic equilibrium equation. 
We used this approach to generate mass ``data-points'' corresponding to each of our 
annuli.  We discuss this 
method, and the models we used to fit our data, in detail in 
\citet{gastaldello07a}. 

For our present analysis, we fitted the density profile with a standard beta-model
functional form, with an added central cusp. To fit the complicated temperature profile,
we adopted a model comprising two smoothly-joined powerlaws. We show the fits to each
profile in Figs~\ref{fig_temp_profile} and \ref{fig_rho_profile}, and the mass data-points
are overlaid on our best-fitting mass model in Fig~\ref{fig_mass_profile}. Clearly, there
is excellent agreement between these two profiles, indicating the robustness of 
our mass analysis results. There are, nonetheless, slight differences if we fit the mass 
data-points directly with our parameterized mass model (we omitted the gas mass, which is 
negligibly small over the range of interest), 
in large part because our adopted temperature profile does not fully capture the 
curvature of the data around 1--2~kpc. Nonetheless, from these data we found
\mbh=$(4.1\pm2.2)\times 10^9$\msun, which, although slightly higher than our best estimate, is 
statistically consistent with it. It is also within 1.6-$\sigma$ of the stellar kinematical
estimate of \citet{gebhardt03a}. 
Similarly, our measured M/\lj\ ratio, $1.24\pm0.22 M_\odot L_\odot^{-1}$
was slightly discrepant, but statistically consistent, with our preferred value.

\section{Systematic error budget} \label{sect_systematics}
\begin{deluxetable}{lll}
\tablecaption{Systematic error budget \label{table_syserr}}
\tablehead{ \colhead{Test} & \colhead{\mbh} & \colhead{M/\lj} \\
\colhead{} & \colhead{($10^9$\msun)} & \colhead{(\mpl)}
}
\tablewidth{3in}
\startdata
Best-fit & 3.35 & 1.37 \\
$\Delta$Statistical & $^{+0.67}_{-0.95}$ & $\pm 0.10$\\ \hline
$\Delta$Method & +0.75 & -0.13 \\
$\Delta$Distance & $^{+0.01}_{-0.27}$& $^{+0.06}_{-0.12}$\\
$\Delta$Plasma code & +0.11 & -0.07 \\
$\Delta$Bandwidth & -0.61 & +0.06 \\
$\Delta$XRB & 0.005& 0.001\\
$\Delta$\nh & 0.001& 0.001 \\
$\Delta$Background & -0.10 & +0.22\\
$\Delta$Centroiding& 0.001 & 0.02\\
$\Delta$Asymmetry & +0.16 & -0.002
\enddata
\tablecomments{Summary of the error-budget for the mass analysis. We list the
best-fitting values of \mbh\ and mass-to-light (M/\lj) and the (90\%) statistical errors
($\Delta$Statistical). In addition we show estimates of the magnitude of the 
systematic uncertainties due to various factors. These should not be added
in quadrature with the statistical errors. We show
the impact of adopting different fitting methods
($\Delta$Method; see text), the estimated errors due to 
distance errors ($\Delta$Distance), 
uncertainties
in the plasma code ($\Delta$Plasma code), the choice of bandwidth 
($\Delta$Bandwidth), 
modelling of unresolved X-ray binaries ($\Delta$XRB), the adopted \nh ($\Delta$\nh),
treatment of the background 
($\Delta$Background), centroiding errors ($\Delta$Centroiding), and
possible asymmetry in the data ($\Delta$Asymmetry). }
\end{deluxetable}

In this section we address the sensitivity of our results to various
data-analysis choices which were made. We focused on those systematic
effects we estimate to be most likely to affect our results.
An estimated upper limit
on the sensitivity of our results to these choices is shown in
Table~\ref{table_syserr}. These numbers reflect the
sensitivity in the best-fit parameter to
each potential source of systematic error, and we stress they
should {\em not} be added in quadrature with the statistical
errors. We outline below how each test was performed. 
Those readers uninterested in the technical details of our analysis may
wish to proceed directly to \S~\ref{sect_discussion}.

First of all, in Table~\ref{table_syserr}, we list ($\Delta$Method) the impact 
of using the traditional mass analysis method described in \S~\ref{sect_nonparametric},
as opposed to the default method we adopted. The traditional method is limited in that
the adopted parameterized models have difficulty in capturing the fine details of the 
temperature and density profiles. Nonetheless, the results are in good agreement from
both techniques. Next  we investigated the impact of changing the distance 
by $\pm$1.8~Mpc, corresponding approximately to the $90$\%\ statistical error 
the I-band SBF distance measurement of \citet{tonry01}. This range encompasses
the adopted distance (16.8~Mpc) used by \citet{gebhardt03a}. This had only a minimal
impact on the measured \mbh, although more effect on the M/\lj\ ratio.

We next considered whether our spectral-fitting choices could have led to a systematic
mis-estimation of the density and temperature, and hence \mbh. There are some 
uncertainties in the modelling of individual emission lines, in particular for the
important Fe blend, so we therefore experimented with replacing the APEC plasma 
model with the MEKAL code, which models the atomic physics differently. This had
only a slight impact on the best-fit results ($\Delta$Plasma code). To assess whether
the adopted bandwidth can affect our results, we experimented with fitting the data in the 
ranges 0.7--7.0, 0.5--2.0~keV and 0.4--7.0~keV, in addition to the default 0.5--7.0~keV.
We estimated the impact on our results of improperly modelling the unresolved X-ray binary
spectral component by varying the temperature of the parameterized (bremsstrahlung) model
by $\pm$25\%, or replacing it with a powerlaw with $\Gamma$=1.5. Finally, to determine
whether systematic errors in the adopted absorbing column could lead to a large error
on \mbh, we perturbed \nh\ by $\pm$25\%\ from its nominal value \citep{dickey90}.

A potentially serious cause for concern when performing spectral analysis of diffuse
emission is the treatment of the background. We therefore investigated whether our
choice for the background made an impact on our results by adopting, instead of 
our preferred treatment given in \S~\ref{sect_spectra}, spectra extracted from
appropriately projected background template events files distributed with the \caldb.
These template files are constructed from processed, stacked observations of
nominally blank-sky fields. Although they do not directly take into account  the 
large-scale variation in the  cosmic X-ray background (nor variability in the 
non X-ray background component), they are widely used to estimate the
background in \chandra\ analysis.
The templates were renormalized to ensure the 9.5--12~keV count-rates of the source
and the background agreed; this is formally incorrect since it also renormalizes the 
cosmic X-ray background in the templates as well. Nonetheless, we found that the 
choice of background did not have a significant impact on \mbh, although the systematic
effect on M/\lj\ is larger than the statistical errors. This is easily understood since 
\mbh\ is determined by data from the bright, central regions of the galaxy, where the
background is negligible.

Another possible cause of systematic error in our analysis is improper centroiding; if our
spectral extraction regions were offset from the dynamical centre of the galaxy, this will
cause the derived temperature and density profiles to be systematically incorrect,
leading to an error in the computation of the central mass. In practice, we would expect
a mis-alignment to result in a systematically under-estimated central temperature, reducing the
significance of the SMBH detection. We checked the agreement of the
\chandra\ centroid and the optical centre (identified by eye from the HST images);
based on a bootstrap alignment of the \chandra\ and HST images, which required a fraction
of the point sources in each to match \citep{humphrey06b}, we estimated they are consistent within 
$\sim$0.6\arcsec, which is comparable to the on-axis PSF of \chandra.
We found that moving the X-ray centre by this amount made little difference to our 
overall result ($\Delta$Centroiding), which is unsurprising since the shift is very much
smaller than the radius of the central extraction region (2.5\arcsec).

Finally, we considered whether low-level asymmetries in the X-ray image might constitute
a strong deviation from our  assumption of spherical symmetry. To investigate the possible
impact of such features, we artificially restricted the field
of view, re-extracting all spectra from a set of suitably-oriented concentric semi-annuli.
We found the fit results agreed very well with our best-fit, indicating that any such asymmetries,
if present, only have a minor impact. It is worth noting that,
if the system globally deviated strongly 
from hydrostatic equilibrium, it would be highly unlikely
that we would observe such symmetry.

\section{Discussion} \label{sect_discussion}
Using the assumptions of hydrostatic equilibrium, spherical symmetry, a single-phase ISM and 
that the stellar mass follows the optical 
light, we have obtained an estimate for the mass of the supermassive
black hole at the centre of \src\footnote{We note that, technically we also require the entropy profile not to behave pathologically
at small radii. However, since hydrostatic equilbrium requires dS/dr $>0$, 
the most discrepant behaviour we might expect to see in S is 
a precipitous drop. In such a case, this would tend to {\em reduce} a central peak in the
model temperature profile, making the
evidence for a supermassive black hole even stronger.}.
This is the first time that the mass of a supermassive black hole has been determined from
the distribution of hydrostatic, X-ray emitting gas. 
In principle, this method can also be applied to other giant elliptical galaxies with
relaxed X-ray morphologies, allowing the most massive black holes to be studied.

We placed a lower limit on the black hole mass of $1.6\times 10^9$\msun, at 99\%\ confidence.
Our measured mass agrees (within $\sim$2-$\sigma$) with that determined from stellar kinematical studies,
$(2.0^{+0.4}_{-0.6})\times 10^9$\msun\ (1-$\sigma$ error), based on a different set of 
assumptions. This makes \src\ one of only a handful of galaxies for which \mbh\ has been measured
{\em via} multiple complementary methods. Although the overall agreement in the shape of 
the \mbh-\sigmac\ relation when measured using different techniques 
\citep[\eg][]{gebhardt00c} provides
good evidence that the masses from stellar kinematics are not {\em on average}
in error, the systematic uncertainties on any individual measurement of \mbh\ are more difficult to
assess. In the few systems where the mass can be inferred from both gas and stellar
dynamics, the results have been mixed, possibly on account of non-circular gas motions 
(\S~\ref{sect_introduction}). Therefore, the good agreement we found between 
the measurements of \mbh\ from stellar dynamics and hydrostatic gas analysis
provides crucial, independent
verification that stellar dynamics can be used to determine reliable SMBH masses, and effectively
rules out pathological orbital structure for the central part of \src.

Our results provide strong support not only for the stellar kinematical results but also the
single-phase, 
hydrostatic equilibrium approximation we adopted. Firstly, the X-ray image shows a remarkably symmetrical 
morphology; we do not confirm the 
evidence of significant asymmetries previously reported with shallower data
\citep{randall03,shurkin07a}. Secondly, the 
hydrostatic mass model actually fits the data remarkably well, despite the complicated temperature
and density profiles. Thirdly, we find good agreement between \mbh\ determined {\em via} both
optical and X-ray techniques.  As our derived mass profile was extended to larger
radii, we found excellent agreement with stellar mass profile 
within $\sim$3~kpc. At even larger radii, the mass determination from the kinematics of 
globular clusters agreed very well with our results \citep{bridges06a}. 
Taken together,
these indicate that the ISM \src\ must be very close to hydrostatic equilibrium at 
all scales, or there is a remarkable conspiracy in the data.

In a recent provocative paper, \citet{diehl07a} analysed the X-ray morphology of a sample of 
early-type galaxies and concluded that, since the majority show clear evidence of morphological
disturbances in their cores, probably all elliptical galaxies are out of hydrostatic equilbrium. 
Quite apart from the obviously false syllogism that, since there exists disturbed (non-hydrostatic)
ISM in many early-type galaxies, {\em all} early-type galaxies (including those without
clear disturbances) are non-hydrostatic, \src\ represents a clear counter-example to this argument. 
This is unsurprising, since  cosmological simulations of galaxy clusters have
shown that the existence of asymmetries typically does not translate to
mass errors of more than 25\%  \citep{tsai94a,evrard96a,nagai07a}, even in systems that are
manifestly more disturbed than \src. Comparing optical and X-ray
mass estimates for two systems (NGC\thin 1399 and M87) that are much more
disturbed than \src, a recent study by \citet{churazov07a}
reaches a similar conclusion. Hence, even when low-level asymmetrical
features are eventually detected with higher quality data, as they must,
there is no evidence from either theoretical or observational studies
that this implies a systematic error in the derived mass much larger than the other
systematic errors we addressed in this paper.


With \chandra\ we were unable to resolve the sphere of influence ($\sim$60~pc) of the SMBH on the 
gas although,
since the emissivity is proportional to the square of the gas density (which rises steeply towards 
the centre), we were able to probe the gas
density within a factor $\sim$2--3 of it. Provided hydrostatic equilibrium holds, resolving
the sphere of influence is not strictly necessary to place constraints on the central mass if
one takes care to compute reliable average densities and temperatures within each radial bin. 
Ideally, however, X-ray imaging spectroscopy at scales smaller than this scale is 
desirable so that the SMBH mass is not determined primarily by a single data-point. With such 
data one could also search for morphological disturbances at the smallest scales, which would 
indicate local deviations from hydrostatic equilbrium. Unfortunately, such imaging is beyond the 
capabilities of \chandra\ and will have to wait until
the era of higher spatial resolution X-ray instruments.
Nonetheless, within the central $\sim$200~pc, the dynamical time-scale is only $8\times 10^5$~yr, 
as compared to a cooling time $\sim 2\times 10^7$~yr so that, unless the gas is being constantly 
stirred up, we would expect it to relax quickly into a state approximating hydrostatic equilibrium. 
Although ISM disturbances due to AGN activity appear common in early-type galaxies 
\citep[\eg][]{jones02a}, these are typically seen on large scales, comparable to any radio
lobes \citep[which, although very weak in \src, are on $\sim$kpc scales:][]{stanger86a}.
Given that the dynamical time-scale is shortest in the 
central parts of the galaxy, \src\ would have to be in a very unusual state for AGN activity
to stir up the gas within only the central $\sim$100~pc
but leave the X-ray morphology outside this range undisturbed. Although the current X-ray data
do not allow us completely to rule this out, turbulence or bulk motion would provide
additional support against gravity \citep[\eg][]{nagai07a}, lowering the central temperature
for a given mass and entropy, and leading to an underestimate of \mbh. Since a deviation from
this assumption would lead to poorer agreement with the optical data, our results indicate that
the gas is very close to hydrostatic.

In addition to being close to hydrostatic equilibrium, our results indicate that 
the gas must also be close to single phase. We tested this explicitly by adding an
extra {\bf vapec} spectral component in each annulus. We found that the fit did not improve
significantly ($\Delta \chi^2$=9 for 30 fewer degrees of freedom) and, excepting the 
innermost bin, the extra component was negligible in each annulus, supporting the single-phase
assumption. Although we cannot rule
out multi-phase gas within the innermost $\sim$200~pc, it is difficult to understand why genuinely
multi-phase gas would be confined only to this region. Fitting a single temperature model
to data where multi-temperature models are actually required in general should lead us to 
underestimate the mass \citep[\eg][]{allen98}, so, if anything, the effect of multi-phase 
gas would be to make our measured \mbh\ more discrepant with the optical data (implying
the gas is probably single-phase). It is worth noting that our models do indicate a 
steeply-rising temperature profile within the innermost bin and in such cases, two-temperature
spectral models are often statistically better descriptions of the data than single-temperature
models \citep[\eg][]{buote00c}. However, one can
still obtain reliable mass profiles by interpreting a single-temperature fit to the data 
provided one takes care to compare its parameters to a suitably averaged model,
as we did in the present work \citep{gastaldello07a}.

One of the intriguing results from our analysis is an explanation of the central temperature spike
seen  in this system. As we show in Fig~\ref{fig_temp_profile}, if we omit the central black hole,
our model temperature profile is much less centrally-peaked than the real data. This confirms the 
predictions of \citet{brighenti99c}, who suggested that such features should be prevalent
when \chandra\ temperature profiles of elliptical galaxies hosting massive black holes are 
investigated. This detection, therefore, opens the possibility of using the same technique to
measure the central black hole mass in other, morphologically relaxed, early-type galaxies.
Since significant X-ray halos are most often associated with the most massive galaxies, 
this will make such a method particularly sensitive to the most massive black holes. 
However, even for a black hole as massive as 
$\sim 3\times 10^9$\msun,
the sphere of influence is only $\sim$50~pc, requiring precise temperature constraints on
a comparable scale. Within the limitations of current instrumentation, this effectively restricts the analysis 
to very X-ray bright galaxies within $\sim$20~Mpc, severely limiting the number of systems in which we 
might hope to see it. With the  improved spatial 
resolution promised by proposed future  X-ray satellites,  such as the \genx\ 
mission \citep{windhorst06a}, though, such measurements should become routine.


In practice, the presence of a central temperature spike is not actually a sufficient diagnostic
for a massive central black hole. As is clear from Fig~\ref{fig_temp_profile}, a mild central 
increase in the temperature can arise as a consequence of the centrally-peaked stellar mass
distribution. In fact, if the stellar
mass was significantly more cuspy (for example in a ``powerlaw'' galaxy, rather than
the ``cored'' \src), we would have seen a significant central temperature rise
even without a black hole.  More importantly, the mass profile alone does not determine the shape of the 
temperature profile, but instead it is the interplay of the mass and entropy profiles. The shape
of the latter is set by the balance between heating and cooling, which is not completely understood.
In general, the flatter the central entropy profile, the more centrally-peaked is the temperature
profile, so quantitative differences in the thermal histories of individual galaxies would explain in part the 
paucity of such temperature spikes. In many objects, this is compounded by the presence of 
AGN-driven disturbances \citep[\eg][]{birzan04a,osullivan05a}, which tend to stir up the gas,
disturbing hydrostatic equilibrium and mixing hot, central gas with cooler gas from larger radii.
Combined with the observational difficulty of the measurement, these effects may explain the 
few reported galaxies exhibiting this feature \citep[\eg][]{humphrey06a,humphrey04b}, in contrast
to the ubiquity of black holes.

In addition to the black hole mass estimate, our method also allowed us to determine 
directly the total stellar mass of the system, without making any assumptions regarding
the stellar mass-to-light ratio, and requiring only that stellar mass follows light.
We therefore obtained {\em both} data-points (albeit not completely independently)
to place on a \mbh\ {\em  versus} stellar mass relation.
We estimated that the mass of the SMBH is $\sim$1.7\%\ times the 
mass in stars of the galaxy. It is interesting to compare this to the relations between \mbh\ 
and bulge mass (\mbul) found by \citet{marconi03a}. For a bulge mass of 
$\sim 2\times 10^{11}$\msun, those authors would predict a ratio of \mbh\ to \mbul\ of 
$\sim$0.2\%, which is discrepant at almost an order of magnitude from our measurement (larger
than the expected scatter).
\citeauthor{marconi03a} estimated \mbul\ by inserting the stellar velocity dispersion and 
effective radius into the Virial relation, which actually gives
an estimate of the {\em total} gravitating matter within, roughly, the effective radius. For the
case of \src, which they include in their sample, they adopted an effective radius of 7.5~kpc (corrected to our adopted distance) and derived 
\mbul $\simeq7.8\times 10^{11}$\msun. This is considerably higher than
the {\em stellar} mass we measured within the galaxy and, not unexpectedly, falls far short of the 
$\sim3\times 10^{13}$\msun\ Virial mass of the entire system \citep{humphrey06a}.
Instead, this is approximately the total gravitating mass
within $\sim$19~kpc, of which dark matter comprises almost $\sim$75\%. Clearly this illustrates the 
difficulty in interpreting Virial theorem estimates for \mbul, especially at the most massive end
where galaxies sit in the centres of group-like dark matter halos. 
In order to gain insight into the SMBH-bulge relation, a full decomposition of the mass profile into
stellar and dark components is, therefore, essential.

Finally, based on the mean stellar population age and abundance estimated from our Lick index analysis
\citep{humphrey06a},
we can estimate the expected mass-to-light (M/\lj) ratio for different simple stellar population
models, assuming all stars were born in a swift burst of star-formation. Using the models
of \citet{maraston05a}, the J-band ratio is $1.86\pm0.21$\mpl, assuming a \citet{kroupa01a} IMF,
and depending only very weakly on the metallicity of the stars.
As an alternative, if we adopt version 2 of the PEGASE code \citep{fioc97a}, the predicted
J-band M/L ratio is $1.55\pm0.16$\mpl, indicating the level of systematic uncertainty in such 
population models.
These results compare favourably with our observed stellar M/\lj\ ratio ($1.37\pm 0.10$\mpl), although
the \citeauthor{maraston05a} models are slightly discrepant.
We note that our measured M/\lj\ is completely inconsistent with the 
predictions for a Salpeter IMF ($2.7\pm0.3$\mpl\ for both sets of models), as we
found in \citet{humphrey06a}, unless the age for the stellar population is dramatically
over-estimated. In that case it would need to be closer to $\sim$7~Gyr than the 13$\pm$2~Gyr we measured.

\acknowledgements
We would like to thank Misty Bentz for useful discussions.
Some of the data presented in this paper were obtained from the Multimission Archive at
the Space Telescope Science Institute ({MAST}).
STScI is operated by the Association of Universities for Research in Astronomy, Inc.,
under NASA contract NAS5-26555.
This research has also made use of the
NASA/IPAC Extragalactic Database (\ned)
which is operated by the Jet Propulsion Laboratory, California Institute of
Technology, under contract with NASA.
Partial support for this work was provided by NASA under 
grant NNG04GE76G issued through the Office of Space Sciences Long-Term
Space Astrophysics Program. Partial support was also provided 
by NASA through Chandra Award Number G07-8083X issued by the Chandra X-Ray Center, which is 
operated by the Smithsonian Astrophysical Observatory for and on behalf of NASA

\bibliographystyle{apj_hyper}
\bibliography{paper_bibliography.bib}

\end{document}